\documentclass{ws-p8-50x6-00}

\def\be{\begin{equation}}
\def\ee{\end{equation}}
\def\bea{\begin{eqnarray}}               
\def\eea{\end{eqnarray}}

\begin{document}

\title{ 
\hspace{7.7cm}
TTP01-30\\ \hspace{7.3cm}ZU-TH-41/01\\
\bigskip
\bigskip
Counting contact terms in  $B\to V\gamma$ decays}

\author{Alexander Khodjamirian~\footnote{\tiny on leave from Yerevan
      Physics Institute, 375036 Yerevan, Armenia}}

\address{Institut f\"ur Theoretische Teilchenphysik,
Universit\"at Karlsruhe, \\D-76128  Karlsruhe, Germany\\ 
E-mail: khodjam@particle.uni-karlsruhe.de}

\author{Daniel Wyler}

\address{Institut f\"ur Theoretische Physik, Universit\"at Z\"urich, 8057
  Z\"urich, Switzerland\\ E-mail:wyler@physik.unizh.ch}


\maketitle

\abstracts{ We clarify the origin and cancellation
of contact terms in the weak 
annihilation amplitudes contributing to $B\to V \gamma$. 
It is demonstrated that the photon emission from the final-state quarks 
vanishes in the chiral limit of massless quarks. 
The contact terms in the QCD light-cone sum rule evaluation of 
the weak annihilation amplitudes are also discussed.}

\begin{center}
{\it to be published in Sergei Matinian Festschrift
``From Integrable Models to Gauge Theories.'', Eds. V.~Gurzadyan, A.~Sedrakyan, World Scientific, 2002}
\end{center}

\section{Introduction}
Radiative decays of $B$-mesons, such as $B \to K^* \gamma$ or 
$B \to \rho \gamma$ provide important tests of the Standard Model and of
new physics scenarios. However, besides the dominant pointlike 
$b \to s \gamma$  or $b \to d \gamma$ 
transitions generated by loops of heavy particles, there 
are ordinary weak decay mechanisms. In particular
the decay $B \to \rho \gamma$ can proceed via the usual 
four-Fermi weak transition accompanied by a photon 
radiated from the quarks inside the initial or final mesons.
Clearly, this ``weak annihilation'' mechanism depicted in Fig.~1
must be under theoretical control in order to predict the 
decay rate with a reasonable accuracy. 

In recent years, several calculations of the weak annihilation 
contribution  to $B\to V \gamma$ were reported
\cite{Yan}$^-$\cite{BMNS}. 
Various techniques, from quark models and  
effective Lagrangians to dispersion relations and 
QCD sum rules were used. In all these calculational schemes 
perturbative photon emission is involved 
which gives rise to gauge non-invariant terms in the photon field,
the so called contact terms.
Clearly these have to vanish in the final answer.

Going through the papers including our own
\cite{KSW} we have to admit 
that the way of handling the contact terms often looks 
arbitrary or even mysterious, and may cause dissatisfaction
among careful readers. In the most recent work \cite{BMNS}, the contact 
terms seem to combine into a gauge-invariant combination and 
contribute to the  physical decay amplitude.

In this paper we would like to investigate this issue 
in more details concentrating on the emergence and cancellation 
of contact terms in the weak annihilation amplitudes in $B \to V 
\gamma$. The elements of our analysis such as 
Ward identities for the 
divergences of conserved currents or decomposition in invariant 
amplitudes are familiar
and were used in some of the papers \cite{GP,BMNS}. However
we are not aware of a complete and comprehensive analysis
which puts all dots over i's. 

\begin{figure}[t]
\epsfxsize=30pc
\epsfbox{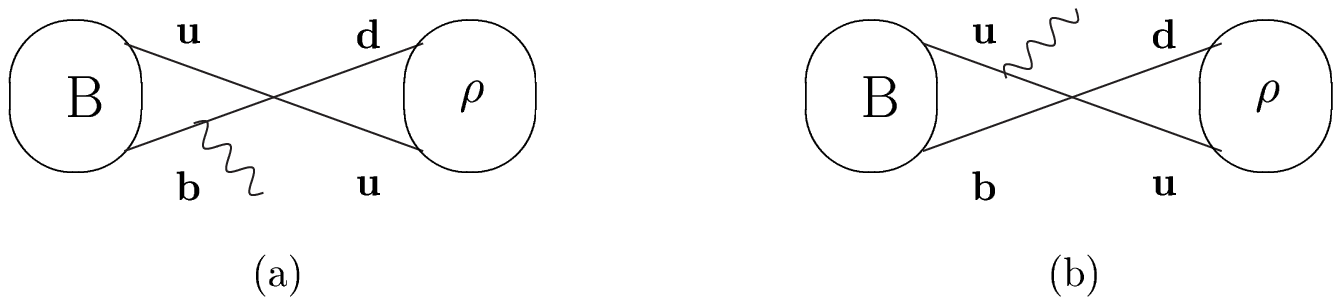} 
\vspace{-2.0cm}
\caption{Weak annihilation mechanism for the decay 
$B ^- \rightarrow \rho^- \gamma $ with photon
emission from the initial state quarks.}
\end{figure}

\section{Radiative leptonic decay $B\to l \nu \gamma$ }

In order to make things as clear as possible, we start our discussion with the
better known decay $B\to l \nu \gamma$ \cite{xuxi}$^-$\cite{KPY} whose 
discussion parallels 
the well known process $\pi \to e \nu \gamma$. It
is customary to divide the amplitude of the latter decay 
into  ``internal bremsstrahlung'' (IB) and 
``structure dependent''(SD) contributions 
(for a review of $\pi \to e \nu \gamma$ see e.g. \cite{pilnugamma}). 
The IB contribution collects the photon 
radiation from the charged lepton (Fig.2~a) and from the meson (Fig.~2b)  
and shows the same helicity (mass) suppression as the 
leading decay $\pi \to l \nu$. However,
a simple calculation of the emission from the lepton yields
a helicity unsupressed amplitude,
\be
A(\pi \to l \nu \gamma)_{IB}=
-ie\frac{G_F}{\sqrt{2}}\bar{u}_l\Gamma_\alpha v_\nu\epsilon^\alpha f_\pi\,,
\label{pi}
\ee
even if $m_l=0$. Here, $e=\sqrt{4\pi\alpha_{em}}$ is the basic electric
charge, $G_F$ the Fermi constant, 
$\Gamma_\alpha=\gamma_\alpha(1-\gamma_5) $, $u_l, v_\nu$ are the lepton
spinors, $\epsilon$ is the 
photon polarization vector and $f_\pi$ the pion decay constant. 
Besides having no helicity suppression, this result is 
also not gauge-invariant in the photon field. But
it is well known that Eq.~(\ref{pi}) is not 
the complete answer for the physical decay amplitude. To recover 
the latter, one has to add the contribution 
of an additional diagram (Fig. 2c) corresponding to an effective 
four-particle
vertex of the form $\pi A_\mu \bar{l} \Gamma^\mu \nu_l$ where $A_\mu$ is
the photon field. 
In the framework of an effective meson theory with a 
point-like pion this vertex can be incorporated
into the usual coupling by replacing the derivative of the pion 
field by the covariant one;  the effective interaction is then
$D_\mu \pi \bar l \Gamma^\mu \nu_l$, where $D_\mu = \partial_\mu -ieA_\mu $. 
Note that the extra interaction is a {\em contact term} in the sense that
the weak and the electromagnetic interactions (currents) originate
in same space-time point. 
\begin{figure}[t]
\epsfxsize=30pc
\epsfbox{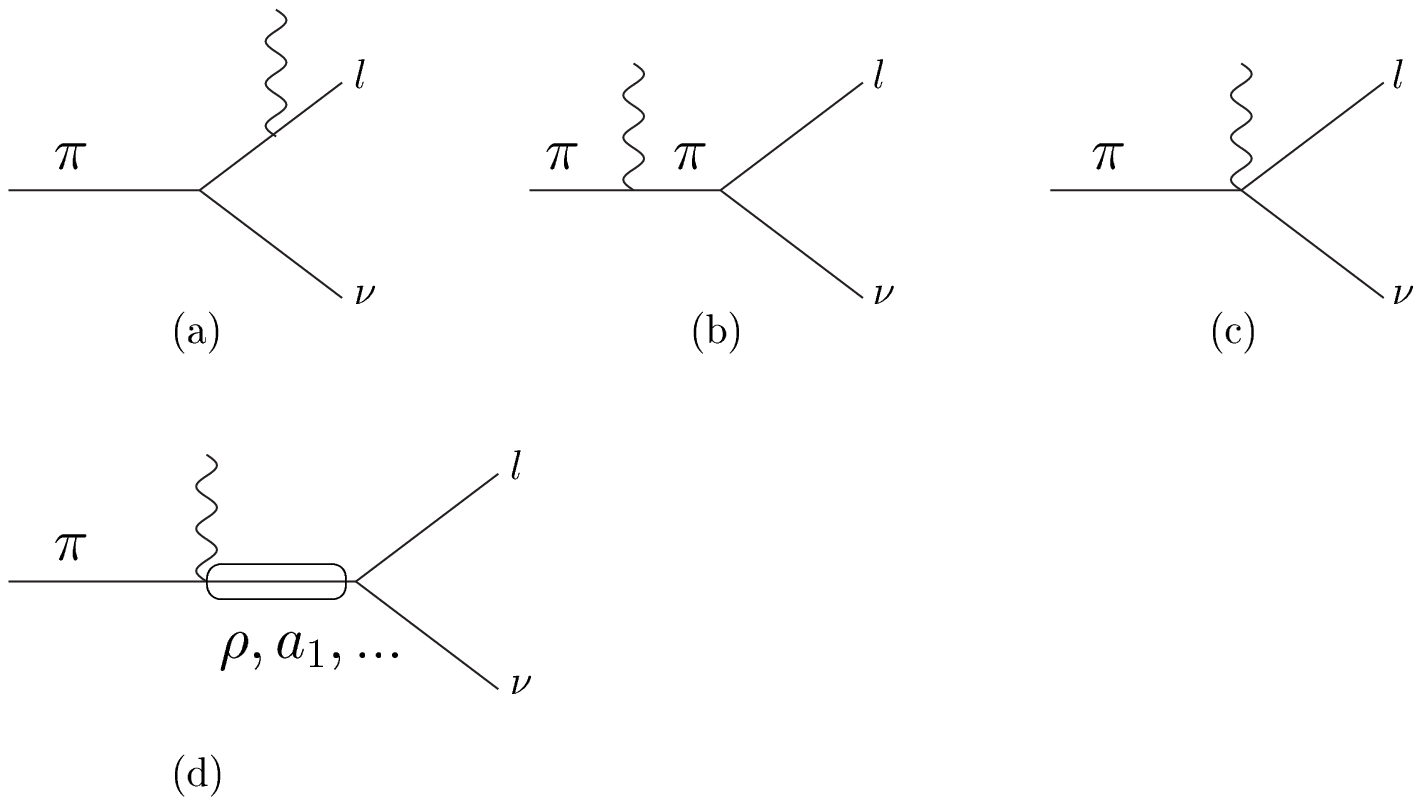} 
\vspace{-1.0cm}
\caption{Diagrams for 
$\pi  \rightarrow l \nu  \gamma $ }
\end{figure}
Fig.~2c yields  
\be
A(\pi \to l \nu \gamma)_{CT}=
ie\frac{G_F}{\sqrt{2}}\bar{u}_l\Gamma_\alpha v_\nu\epsilon^\alpha f_\pi\,,
\label{piCT}
\ee
exactly canceling the amplitude (\ref{pi}):
The emission of the photon from 
the final state leptons is indeed absent in the ``chiral limit'' of 
massless leptons.
The only surviving contribution comes from the photon emission
in the initial state which is ``structure dependent'' 
and corresponds to the diagrams of Fig. 2d. The corresponding 
amplitude can be parametrized in a gauge-invariant form
\bea
A(\pi \to l \nu \gamma)_{SD} = 
e\frac{G_F}{\sqrt{2}}\left(\bar{u}_l \Gamma_\alpha v_\nu\right)\Big(
iF^{(\pi)}_A(p^2)( \epsilon^\alpha(p\cdot q)-(\epsilon\cdot p) q^{\alpha})
\nonumber
\\
+ 
F^{(\pi)}_{V}(p^2) \epsilon^{\alpha\mu\lambda\rho}\epsilon_\mu
p_\lambda q_\rho \Big).
\label{sdpi}
\eea
Here, $p$ is the sum of the lepton momenta and $q$ is the momentum 
of the real photon, $q^2=0$.
The SD amplitude  is determined by the axial 
and vector form factors $F^{(\pi)}_{A,V}$.

The SD contributions come from  
intermediate quark-antiquark states with $J^P=1^-$ and 
$1^+$ and are therefore very difficult to calculate without a 
reliable method to handle long-distance quark-gluon interactions.
A way out is to represent the form factors $F_{A,V}^{(\pi)}$ in terms of 
dispersion relations in the variable $p^2$ with a set of 
intermediate hadronic states with the quantum numbers of 
vector and axial-vector mesons, respectively. 

We now consider the  radiative leptonic decays in the 
framework of $QCD$ where mesons are composite particles. The amplitude 
for $B\to l\nu \gamma$ can be written as 
\be
A(B^-\to l\bar{\nu}_l \gamma) = \frac{G_F}{\sqrt{2}}V_{ub}
\langle\, l \bar{\nu}(p)~ \gamma(q) \mid 
\left(\bar{l}\Gamma_\rho \nu \right)
\left(\bar{u}\Gamma^\rho b \right)\mid B^-(p+q)\,\rangle\,.
\label{def1}
\ee
To first order in the e.m. interactions the matrix element above 
can be rewritten as a sum of two physically distinct contributions: 
\bea
&&
\langle\, l \bar{\nu}(p)~ \gamma(q) \mid 
\left(\bar{l}\Gamma_\rho \nu \right)
\left(\bar{u}\Gamma^\rho b \right)\mid B^-(p+q)\,\rangle
\nonumber
\\
&&=ie\epsilon^\mu \Big[ (\bar{u}_l\Gamma_\rho v_\nu)
\!\int \!d^4x\,e^{iqx} 
\langle 0 \mid T\{ j_\mu^{em}(x)\,\bar{u}\Gamma^\rho b(0) 
\}\mid B^-(p+q)\rangle 
\nonumber
\\
&&-\!\int \!d^4x \,e^{iqx}\langle l \bar{\nu}(p) \mid T\{ j_\mu^{em}(x) 
\bar{l}\Gamma_\rho \nu(0) \}\mid 0\rangle if_B(p+q)^\rho\Big]\,,
\label{blnugamma}
\eea
where $j_\mu^{em}= -\bar{l}\gamma_\mu l + \sum_{q=u,d,s,c,b} 
e_q\bar{q}\gamma_\mu q$ is the e.m. current and $e_q$ the quark
e.m. charges in the units of $e$.
The first term on the r.h.s. of Eq.~(\ref{blnugamma})
corresponds to the photon emission from the initial $B$ meson state
and the leptonic part is trivially factorized out.
In the second term the photon is emitted from the final 
charged lepton and the hadronic matrix element 
is factorized using the standard 
definition of the B-meson decay constant :
$\langle 0 \mid\bar{u}\gamma^\rho\gamma_5 b\mid B(p+q)\rangle=  
if_B(p+q)^\rho$.
The remaining lepton-photon matrix element in this term 
can be simply calculated
using Feynman rules of QED. As expected, at $m_l=0$  the result is 
very similar to Eq.~(\ref{pi}):
\be
e\epsilon^\mu\!\!\int\! d^4x \,e^{iqx}\langle l \bar{\nu}(p) \mid T\{ j_\mu^{em}(x) 
\,\bar{l}\Gamma_\rho \nu (0)\}\mid 0\rangle f_B(p+q)^\rho \!= 
\!-ie\bar{u_l}\Gamma_\mu v_\nu \epsilon^\mu f_B \,, 
\label{ctBl}
\ee
and again has a typical structure of a contact, gauge non-invariant term. 
In analogy to the pion case 
we expect that photon emission from the initial $B$ meson contains a 
contact term which cancels Eq.~(\ref{ctBl}).

To see this explicitly we use a generic 
covariant decomposition of the hadronic matrix element
\be
\label{matrix}
T^{(B)}_{\mu\rho}(p,q) = i\int\! d^4x\, 
e^{iqx}\langle 0 \mid T\{ j_\mu^{em}(x)\,\bar{u}\Gamma_\rho b(0) 
\}\mid B^-(p+q)\rangle\ee
in two independent 4-momenta $p$ and $q$:
\be
T^{(B)}_{\mu\rho}(p,q)= g_{\mu\rho}\,a +p_\mu q_\rho\,b +q_\mu p_\rho\,c
+p_\mu p_\rho\,d +q_\mu q_\rho\,e +
\epsilon_{\rho\mu\lambda\sigma}p^\lambda q^\sigma F^{(B)}_V\,.
\label{decomp}
\ee
where $a,b,c,d,e$ and $F^{(B)}_V$ are invariant amplitudes.
We apply the standard electromagnetic Ward identity to the matrix
element (\ref{matrix}) using the conservation of 
the e.m. current. In momentum space it corresponds 
to a multiplication by $q_\mu$. The well-known additional contribution due to 
differentiation of the $\theta$-function in the $T$ product 
yields a contact term:
\be
q^\mu T^{(B)}_{\mu\rho}= i(p+q)_\rho f_B\,.
\ee 
Applied to the decomposition (\ref{decomp}),
the same operation yields
\be
q^\mu T^{(B)}_{\mu\rho}= q_\rho\,a+(p\cdot q)q_\rho\, b +(p\cdot q)p_\rho\, d\,. 
\ee
Comparing the coefficients in two above equations 
at independent 4-momenta one gets the relations 
\be
a+(p\cdot q)\,b= if_B,
\label{relab}
\ee
and 
\be
(p\cdot q)\,d= if_B,
\label{reld}
\ee
The first of these relations
connects the unknown amplitudes $a$ via $b$ whereas the second 
fixes the amplitude $d$.
As a result we can rewrite $T^{(B)}_{\mu\rho}$  
in the following general form
\bea
\label{arbi}
T^{(B)}_{\mu\rho}(p,q)= (g_{\mu\rho}(p\cdot q) -p_\mu q_\rho)\,
iF_{A}^{(B)}+ g_{\mu \rho}(p\cdot q)\,\alpha +p_\mu q_\rho\,\beta 
\nonumber
\\
+ q_\mu p_\rho\,c  
+i\frac{p_\mu p_\rho}{(p\cdot q)}f_B
+q_\mu q_\rho\,e + 
\epsilon_{\rho\mu\lambda\sigma}p^\lambda q^\sigma\,F_{V}^{(B)}\,,
\label{decompt}
\eea
introducing new invariant amplitudes $F_{A}^{(B)}$,
$\alpha$ and $\beta$ where the latter must satisfy the condition
(Ward identity)\be
\label{cond}
\alpha + \beta =i\frac{f_B}{(p\cdot q)}.
\ee 
The values of $\alpha$ and the 
corresponding $\beta$ themselves are 
arbitrary and not fixed by the electromagnetic Ward identity. 
In Eq.~(\ref{decompt}) the terms proportional to $F_{A}^{(B)}$, 
$c$ and $F_{V}^{(B)}$
are gauge-invariant (they vanish after being multiplied by $q_\mu$) 
whereas the term proportional to $f_B$ disappears in the chiral limit
after being multiplied by the lepton current.  
The remaining contact-term part of $T^{(B)}_{\mu\rho}$ 
containing $\alpha$ and  $\beta$ 
is gauge noninvariant. Different choices of $\alpha$ and $\beta$ simply
reflect different choices of $F_{A}^{(B)}$ and allow us
to rewrite  $T^{(B)}_{\mu\rho}$ in many ways.

Let us set $\beta=0$; this gives 
\bea
T^{(B)}_{\mu\rho}(p,q)= 
(g_{\mu\rho}(p\cdot q) 
-p_\mu q_\rho)\,iF_{A}^{(B)} +ig_{\mu \rho}f_B 
\nonumber
\\
+ q_\mu p_\rho\,c +i\frac{p_\mu p_\rho}{(p\cdot q)}f_B
+q_\mu q_\rho\,e 
+\epsilon_{\rho\mu \lambda\sigma}p^\lambda q^\sigma\,F_{V}^{(B)}.
\label{decomp2}
\eea
Substituting  Eq.~(\ref{decomp2}) together with Eq.~(\ref{ctBl})
in Eq.~(\ref{blnugamma})
one finally obtains for the decay amplitude 
\bea
A(B^-\to l\nu \gamma) = &&e\frac{G_F}{\sqrt{2}}V_{ub}\Big\{
\left(\bar{u_l}\Gamma^\rho v_\nu \right)\Big[
\left(\epsilon_{\rho}(p\cdot q) - (\epsilon\cdot p)\,q_\rho\right)\,iF_{A}^{(B)}
+i\epsilon_\rho\, f_B
\nonumber
\\
&&+\epsilon_{\rho\mu\lambda\sigma}\epsilon^\mu p^\lambda q^\sigma\,F_{V}^{(B)} 
\Big]
\!-i\left(\bar{u_l}\Gamma^\rho v_\nu \right)\epsilon_\rho\, f_B \Big\}\,,
\label{blnug}
\eea
where the terms in the brackets correspond to the initial state photon 
and the remaining term is the 
only effect of the 
final state emission, the contact term (\ref{ctBl}).
We see that the two contact terms cancel in the r.h.s. of this expression 
as expected. The 
remaining ``structure dependent'' amplitude
is a combination of two gauge-invariant
form factors.

The story is however not yet finished. What about choosing another
set of values for $\alpha$ and $\beta$?
Instead of a contact term proportional to $f_B\epsilon_\rho$ 
in  Eqs.~(\ref{decomp2})  and (\ref{blnug}) 
we will recover the combination 
$ \alpha (p \cdot q )\epsilon_\rho +\beta  (p\cdot \epsilon) q^\rho $  
which looks completely different. But when the ``final-state'' contact 
term  Eq.~(\ref{ctBl}) is added and the relation (\ref{cond})
is used, a gauge invariant
result is obtained, with a form factor 
$ iF_{A}^{(B)} \to iF_{A}^{(B)} -\beta$. The choice $\beta\neq 0$ 
looks less attractive, because
it leaves the impression that the helicity unsuppressed 
contact term of the leptonic photon emission is part of the answer.
But it is not necessarily wrong. Everything depends 
on how the 'gauge invariant' form factor $ F_{A}^{(B)}$ is  
calculated in a given framework. If we calculate the 
coefficient of the kinematical structure $g_{\mu\rho}$ in 
$T^{(B)}_{\mu\rho}$, 
then we have to add the additional contact term because 
the diagrams used to calculate this invariant amplitude implicitly 
contain such a term.\footnote{one of us (A.K.) is grateful to D. Melikhov for
a discussion on this point.} Below we will see this in the particular
example of the correlation function used in light-cone sum rules.
We prefer the scheme which corresponds to the $\beta=0$ choice
where the contact terms vanish and which is in accordance 
with the physical picture. Since the structure 
of the contact term in the final state radiation 
is fixed (it is the $g_{\mu\rho}$ structure), one should look for a
formulation of the initial state radiation which yields the {\bf other} 
kinematical
structure (in this case $p_\mu q_\rho$). This statement is clearly 
independent of the choice of $\beta$.

Thus, we have once more convinced ourselves 
that there is no photon emission from charged 
leptons in the massless lepton limit.
It is clear that the same statement will be valid for 
final massless quarks if the strong interactions are neglected, that is 
for the weak annihilation
contribution to the $B\to X_d \gamma$ inclusive width calculated 
at the partonic level. But since the inclusive 
width is a sum of positive exclusive widths, it is 
not possible that {\em any} of the exclusive channels gets chirally 
unsuppressed final state radiation.
Nevertheless, let us demonstrate that explicitly using
the same technique of symmetry relations and Ward identities.

\section{$B\to \rho \gamma$}
We now turn to the case of interest, the decay 
$B^- \to \rho^- \gamma$. In order 
to trace the contact terms we choose the
charged $B$ mode and consider only the weak annihilation
decay mechanism; furthermore we employ the factorization approximation. 
The decay amplitude is similar to Eq.~(\ref{def1}) if the lepton
pair is replaced by the light-quark pair and the final state
correspondingly by a $\rho$ meson:
\be
A(B^-\!\to \!\rho^- \gamma)_{WA} = \frac{G_F}{\sqrt{2}}V_{ub}V^*_{ud}
\langle\,\rho^-(p)~ \gamma(q) \!\mid\! 
\left(\bar{d}\Gamma_\nu u \right)
\left(\bar{u}\Gamma^\nu b \right)\!\mid \!B^-(p+q)\,\rangle 
\label{Abrhogamma}.
\ee 
In the above, we omit the combination of 
Wilson coefficients $a_1=c_1+c_2/3$ which is numerically close 
to 1 and irrelevant for our discussion.
Again, the matrix element can be rewritten as a sum of two
contributions:
\bea
\langle\, \rho(p)~ \gamma(q) \mid 
\left(\bar{d}\Gamma_\nu u \right)
\left(\bar{u}\Gamma^\nu b \right)\mid B^-(p+q)\,\rangle 
=e \epsilon^\mu \epsilon^{(\rho)\nu} f_\rho m_\rho T^{(B)}_{\mu\nu}
\nonumber
\\
-ie\epsilon^\mu (p+q)^\nu f_B\,T^{(\rho)}_{\mu\nu}\,,
\label{brhogamma}
\eea
where $\epsilon^{(\rho)}$ is the polarization vector, 
$\epsilon^{(\rho)}\cdot p=0$ and $f_\rho$ is the decay constant
of $\rho$, defined as 
$\langle \rho(p)\mid \bar{d}\gamma_\nu u \mid 0\rangle= 
m_\rho f_\rho\epsilon^{(\rho)}_\nu$. 
The first term in Eq.~(\ref{brhogamma})
contains  the matrix element of the photon 
emission from the $B$ meson already analysed 
in the previous section. The second represents the 
final-state emission and includes 
the hadronic matrix element 
\be
T^{(\rho)}_{\mu\nu}= 
i \!\int\! d^4x\,e^{iqx} \langle\, \rho(p) \mid T\{ j_\mu^{em}(x)\, 
\bar{d}\Gamma_\nu u(0) \}\mid 0\rangle\, 
\ee
which is multiplied by $p+q$.
To analyse this object we again apply 
the Ward identity for the e.m. current:
\be
q^\mu T^{(\rho)}_{\mu\nu}= f_\rho m_\rho \epsilon^{(\rho)}_\nu\,,
\ee
or equivalently
\be
q^\mu (p+q)^\nu T^{(\rho)}_{\mu\nu}= f_\rho m_\rho(\epsilon^{(\rho)}\cdot q)\,. 
\label{wirho}
\ee

The general decomposition
of the product $(p+q)^\nu T^{(\rho)}_{\mu\nu}$ reads
\bea
(p+q)^\nu T^{(\rho)}_{\mu\nu}= 
\epsilon^{(\rho)}_\mu\,a^{(\rho)} 
+(\epsilon^{(\rho)}\cdot q)p_\mu \,b^{(\rho)}
\nonumber
\\
+ 
(\epsilon^{(\rho)}\cdot q)q_\mu\,c^{(\rho)} 
+\epsilon_{\nu\mu\lambda\sigma}\epsilon^{(\rho)\nu} p^\lambda q^\sigma 
\,F_{V}^{(\rho)}.
\label{decompf}
\eea
Multiplying both parts of this equation by $q_\mu$ one obtains:
\be
q^\mu (p+q)^\nu T^{(\rho)}_{\mu\nu}= 
(\epsilon^{(\rho)}\cdot q)\,a^{(\rho)}
+
(\epsilon^{(\rho)}\cdot q)(p\cdot q)\,b^{(\rho)}.
\label{multi}
\ee
The Ward identity (\ref{wirho}) and Eq.~(\ref{multi})
yield then a relation between the invariant amplitudes $a^{(\rho)}$
and $b^{(\rho)}$ :
\be
a^{(\rho)}+ (p\cdot q)\,b^{(\rho)} = f_\rho m_\rho\,.
\label{emWI}
\ee
It might seem again that there is an arbitrariness 
(as in Eq.~(\ref{arbi})) in writing the amplitude.
However, at this point it is important to notice that 
the final-state weak current is also conserved in the chiral
limit. Therefore there is an additional Ward identity for the 
hadronic matrix 
element  $T^{(\rho)}_{\mu\nu}$:
\be
(p+q)^\nu T^{(\rho)}_{\mu\nu} = f_\rho m_\rho \epsilon^{(\rho)}_\mu\,. 
\ee
The situation is 
simplified even more by the fact that the  
product $(p+q)^\nu T^{(\rho)}_{\mu\nu}$ entering 
the decay amplitude is by itself the l.h.s. of the Ward identity.
The result of this consideration are the constraints
\be
a^{(\rho)} =f_\rho m_\rho,~~ b^{(\rho)} =c^{(\rho)}=F_{V}^{(\rho)}=0\,,
\label{constr}
\ee 
consistent with the e.m. Ward identity (\ref{emWI}). 
Most importantly, Eq.~(\ref{constr}) fixes the final state emission uniquely 
as was the case in the leptonic decay. 
The final expression for the $B\to \rho \gamma$ weak annihilation
amplitude is then
\bea
A(B^-\!\to\! \rho^- \gamma)_{WA}\!&&=\!
e\frac{G_F}{\sqrt{2}}V_{ub}V_{ud}^*
f_\rho m_\rho\Big\{\Big[\Big((\epsilon \cdot \epsilon^{(\rho)})
(p\cdot q)
-(\epsilon\cdot p)(\epsilon^{(\rho)}\cdot q)\Big)\,iF_{A}^{(B)}
\nonumber
\\
&&+if_B(\epsilon \cdot \epsilon^{(\rho)})
+\epsilon_{\nu\mu\lambda\sigma}
\epsilon^{(\rho)\nu}\epsilon^\mu p^\lambda q^\sigma\, F_{V}^{(B)}\Big]
-if_B (\epsilon \cdot \epsilon^{(\rho)})
\Big\}\,,
\label{bgammarho}
\eea
where the part proportional to $T^{(B)}_{\mu\nu}$ is 
indicated by brackets. The   
contact terms  again cancel each other, if, as explained in the 
previous section, 
the structure proportional to $p_\mu q_\nu$
has been chosen to calculate the form factor $F_{A}^{(B)}$.   
Thus
in the chiral limit there is no photon emission from the final $\rho$ 
in $B\to \rho \gamma$. We also note that $ F^{(\rho)}_{V}=0$ 
prohibits the emission also in the vector part of the amplitude
.\footnote{ Sometimes the axial and vector form factors for 
$B\to V \gamma$ are called parity-violating and parity-conserving, 
respectively.} The weak annihilation amplitude again has the form 
of a sum of two ``structure dependent'' terms
corresponding to the photon emission from the 
initial state.

\section{$B\to D^*\gamma$}

We now come to the physically different 
case where the final state contains a $c$ quark, e.g. the $B\to D^* \gamma$
decay which is in fact dominated by the weak annihilation 
mechanism. The point is that the mass of the charm quark can not
be neglected. While a more complete analysis of this decay using
QCD light-cone sum rules will be presented elsewhere  
\cite{BKSW}, we limit ourselves to the issue 
of contact terms. The decay amplitude is similar to 
Eq.~(\ref{Abrhogamma}) with obvious replacements 
in the quark current and final state: 
\be
A(B^-\to D^{*-} \gamma)= \frac{G_F}{\sqrt{2}}V_{ub}V^*_{cd}
\langle\,D^{*-}(p)~ \gamma(q) \mid 
\left(\bar{d}\Gamma_\nu c \right)
\left(\bar{u}\Gamma^\nu b \right)\mid B^-(p+q)\,\rangle 
\label{AbDgamma}\,.
\ee 
In the factorization approximation: 
\bea
&&\langle\, D^*(p)~ \gamma(q) \mid 
\left(\bar{d}\Gamma_\nu c \right)
\left(\bar{u}\Gamma^\nu b \right)\mid B^-(p+q)\,\rangle 
\nonumber
\\
&&=e \epsilon^\mu \epsilon^{(D^*)\nu} f_{D^*} m_{D^*}T^{(B)}_{\mu\nu}
-i e\epsilon^\mu (p+q)^\nu f_B T^{(D^*)}_{\mu\nu},\,
\label{bDgamma}
\eea
where $\epsilon^{(D^*)}$ and $f_{D^*}$ are the polarization vector 
and the decay constant of $D^*$.
The hadronic matrix element $T^{(B)}_{\mu\nu}$ responsible for the
initial-state photon emission is exactly the same as in 
$B\to l\nu \gamma$ or $B\to \rho \gamma$, whereas  
the matrix element determining
the photon emission from the final $D^*$ is defined as
\be
T^{(D^*)}_{\mu\nu}= i \!\int d^4x \,e^{iqx}\langle D^*(p) \mid T\{ j_\mu^{em}(x) \,\bar{d}\Gamma_\nu c(0) \}\mid 0\rangle\,,
\ee
With the most general decomposition
(\ref{decompt}) we obtain for the first term on r.h.s.
of Eq.~(\ref{bDgamma}):
\bea
&&e\epsilon^\mu \epsilon^{(D^*)\nu} f_D^* m_{D^*}T^{(B)}_{\mu\nu}=
 ef_D^*m_{D^*}\Big\{
\left[(\epsilon \cdot \epsilon^{(D^*)})(p\cdot q)-
(\epsilon \cdot p)(\epsilon^{(D^*)}\cdot q)\right]iF_A^{(B)}
\nonumber
\\
&&+(\epsilon\cdot\epsilon^{(D^*)})(p\cdot q)\,\alpha +
(p\cdot \epsilon)(q \cdot \epsilon^{(D^*)})\,\beta 
+\epsilon_{\nu\mu \lambda\sigma}\epsilon^{(D^*)\nu}\epsilon^\mu
p^\lambda q^\sigma\,F_{V}^{(B)} \Big\}\,,
\label{TBD}
\eea
with $\alpha$ and $\beta$ related by Eq.~(\ref{cond}).
The product  $(p+q)^\nu T^{(D^*)}_{\mu\nu}$ for the final state
emission can again be constrained 
by the Ward identity of the weak current.
In this case the $\bar{d}\Gamma_\nu c$ current is {\bf not}
conserved: $\partial_\nu (\bar{d}\Gamma_\nu c) = m_c\bar{d}(1-\gamma_5)c$
and therefore
\bea
(p+q)^\nu T^{(D^*)}_{\mu\nu} &&= f_D^*m_D^* \epsilon^{(D^*)}_\mu
\nonumber
\\
&&+ im_c\int d^4x e^{ipx}\langle D^*(p) \mid T\{ j_\mu^{em}(x) 
\bar{d}(1-\gamma_5) c \}\mid 0\rangle\,.
\eea
Matching this expression with the general decomposition 
analogous to Eq.~(\ref{decompf})
\bea
(p+q)^\nu T^{(D^*)}_{\mu\nu}= 
\epsilon^{(D^*)}_\mu\,a^{(D^*)} 
+(\epsilon^{(D^*)}\cdot q)p_\mu \,b^{(D^*)}
\nonumber
\\
+ 
(\epsilon^{(D^*)}\cdot q)q_\mu\,c^{(D^*)} 
+\epsilon_{\nu\mu\lambda\sigma}\epsilon^{(D^*)\nu}p^\lambda  q^\sigma 
\,F_{V}^{(D^*)}\,,
\label{decompD}
\eea
we conclude that 
\be
a^{(D^*)} = f_{D^*}m_{D^*} + O(m_c),~~~  
b^{(D^*)},c^{(D^*)},F_V^{(D^*)}\sim O(m_c)\,,
\ee
that is the photon emission from the final state 
has an nonvanishing amplitude proportional to $m_c$.
Finally, the electromagnetic Ward identity yields a relation
similar to (\ref{emWI}):
\be
a^{(D^*)}+(p\cdot q)\,b^{(D^*)}= f_{D^*}m_{D^*}\,.
\ee
As in the case of $T^{(B)}$ analysed above there
is a certain  freedom in using this constraint. 
In particular, it is possible to rewrite   
\bea
&&(p+q)^\nu T^{(D^*)}_{\mu\nu}
= \left(\epsilon^{(D^*)}_\mu (p\cdot q) - p_\mu (\epsilon^{(D^*)}
\cdot q)\right)\,iF_{A}^{(D^*)}  
+\epsilon^{(D^*)}_\mu
(p\cdot q)\,\alpha^{(D^*)} 
\nonumber
\\
&&+ (\epsilon^{(D^*)}\cdot q) p_\mu\beta^{(D^*)}
+ (\epsilon^{(D^*)}\cdot q)q_\mu \,c^{(D^*)} 
+\epsilon_{\nu\mu\lambda\sigma}\epsilon^{(D^*)\nu} p^\lambda q^\sigma 
\,F_{V}^{(D^*)}\,,
\label{decompD1}
\eea
introducing new amplitudes $ F_{A}^{(D^*)}$, $\alpha^{(D^*)}$ and 
$\beta^{(D^*)}$ with the condition  
\be
\alpha^{(D^*)} +\beta^{D^*}=\frac{m_{D^*}f_{D^*}}{(p\cdot q)}\,.
\label{cond2}
\ee
Multiplying Eq.~(\ref{decompD1}) by $-ie\epsilon^\mu f_B$ we obtain
the general form of the final-state emission amplitude which 
has to be added to the initial-state one given in Eq.~(\ref{TBD}).
It seems that we have now a problem in adjusting the arbitrary 
coefficients in both the initial and the final state emission. However, 
it is easy to see that the constraints~(\ref{cond}) and (\ref{cond2})
allow to arrange the gauge-invariant combinations 
of form factors for both terms $T^{(B)}$ and $T^{(D^*)}$ {\em
separately}
so that the remaining contact terms proportional to $f_Bf_{D^*}$ cancel.
Like for the initial emission where the choice $\beta=0$ was 
preferable, we also suggest using $\beta^{D^*}=0$ for $T^{(D^*)}$.
In this case the final expression for the decay amplitude 
\bea
A(B^-\to D^{*-} \gamma)&&= e\frac{G_F}{\sqrt{2}}V_{ub}V_{cd}^*
\Bigg\{ 
\Big[\left(
(\epsilon \cdot \epsilon^{(D^*)})(p\cdot q)-
(\epsilon\cdot p)(\epsilon^{(D^*)}\cdot q)  
\right)\,iF_{A}^{(B)}
\nonumber
\\
&&+if_B(\epsilon \cdot \epsilon^{(D^*)})
+\epsilon_{\nu\mu\lambda\sigma}
\epsilon^{(D^*)\nu} \epsilon^\mu
p^\lambda q^\sigma\, F_{V}^{(B)}\Big]
m_{D^*}f_{D^*}
\nonumber
\\
&&-if_B\Big[\left((\epsilon \cdot \epsilon^{(D^*)})(p\cdot q) - 
(\epsilon \cdot p)(\epsilon^{(D^*)}\cdot q)\right)\,iF_{A}^{(D^*)}
\nonumber
\\
&&+f_{D^*}m_{D^*}(\epsilon \cdot \epsilon^{(D^*)})
+i\epsilon_{\nu\mu\lambda\sigma}\epsilon^{(D^*)\nu} \epsilon^\mu 
p^\lambda q^\sigma \,F_{V}^{(D^*)}\Big] \Bigg\}\,,
\eea
contains, in addition to the form factors $F_{A,V}^{(B)}$ of the massless
case, also two new form factors $F_{A,V}^{(D^*)}$ corresponding 
to the photon emission from the $D^*$.  The respective contact terms cancel 
as desired: None of the spurious contact terms appear
in the gauge-invariant amplitudes. As before, the relevant
structure to calculate is $(\epsilon \cdot p)(q\cdot\epsilon^{(D^*)})$. This
concludes our general considerations. 

\section{Applying light-cone sum rules to $B\to V \gamma$}

As pointed out before, the way the structure-dependent contributions are
calculated is important for identifying the right terms and
discarding the ones which cancel. 
Let us illustrate this for the 
QCD light-cone sum rule (LCSR) approach
used to calculate the weak annihilation amplitude for  
$B\to \rho \gamma$ \cite{KSW,AB} and $B\to l \nu \gamma$ \cite{KSW}
previously. In these papers the object $T^{(B)}_{\mu\nu}$ defined in 
Eq.~(\ref{matrix}) was calculated. The method is to
introduce a correlation function
\be
\label{corr}
\Pi_{\nu}(p,q) = i\!\int\! d^4x\, e^{ipx}
\langle 0 \mid T\{ \bar{u}\Gamma_\nu b(x),m_b\bar{b}i\gamma_5 u(0)\}
\mid 0\rangle_{F(q)} 
\ee
of two heavy-light currents in the external photon field with momentum 
$q$. The $B$ meson is interpolated by the quark current.
In first order in the e.m. interaction 
$
\Pi_\nu=e\epsilon^\mu \Pi_{\mu\nu}
$
where
\be
\Pi_{\mu\nu}(p,q)\! = \!\!-\!\!\int \!d^4x\,d^4y \,e^{ipx+iqy}
\langle 0 \mid T\{ j_\mu^{em}(y),\bar{u}\Gamma_\nu b(x),
m_b\bar{b}i\gamma_5 u(0)\}\!\mid 0\rangle. 
\label{pimunu}
\ee
The Ward identity with respect to the e.m. current for this
three-point correlator can be easily derived with the following
nontrivial result:
\be
q^\mu \Pi_{\mu\nu}= p_\nu \Pi^{CT}(p^2) -
(p+q)_\nu\, \Pi^{CT}((p+q)^2)\,,
\label{ctcorr}
\ee
where $\Pi^{CT}$ is the invariant amplitude 
determining the two-point correlator 
\be
\Pi^{CT}_{\nu}(r) = i\!\int\! d^4x\, e^{irx}
\langle 0 \mid T\{ \bar{u}\Gamma_\nu b(x),m_b\bar{b}i\gamma_5 u(0)\}\
\mid 0\rangle = r_\nu \Pi^{CT}(r^2)\,. 
\ee

Before continuing, let us make the following remark.
In the LCSR approach the dominant contribution
to the correlation function comes from the
long-distance photon emission parametrized by
the light-cone distribution amplitudes of the photon.
This part is explicitly gauge-invariant in e.m. field
and cannot give rise to contact terms. Our concern here
is the short-distance part of Eq.~(\ref{corr}) which
corresponds to the perturbative emission of the photon
from quark lines.
                                   
Returning to the counterterms, we see that a combination of the 
correlators in Eq.~(\ref{ctcorr})
now plays the role of the contact term 
for the three-point correlation function.
Importantly, one of them is a function of $p$ only, 
whereas the other one depends on $p+q$. 
Diagrammatically, the first term in Eq.~(\ref{ctcorr})
corresponds to the emission of the photon in the point of the
$B$ meson current whereas the second one in the point of the 
weak current.
The correlation function (\ref{corr}) itself
is calculated using the operator-product expansion
(OPE); the result can be expressed as a general decomposition
\be
\Pi_{\mu\nu}= g_{\mu\nu}\Pi_a + p_\mu q_\nu \Pi_b
+   q_\mu p_\nu \Pi_c + p_\mu p_\nu \Pi_d+ q_\mu q_\nu \Pi_e
+i\epsilon_{\mu\nu\lambda\sigma}p^\lambda q^\sigma\Pi_V \,,
\label{decompPi}
\ee
in terms of independent invariant amplitudes $\Pi_{a}$, $\Pi_{b}$, etc. 
From the Ward identity (\ref{ctcorr})
we see that the result is not explicitly gauge-invariant:
The amplitudes in Eq.~(\ref{decompPi}) do not combine to 
form a gauge-invariant expression. If we multiply both parts of 
Eq.~(\ref{decompPi}) by $q_\mu$ and compare 
the result with Eq.~(\ref{ctcorr}) we obtain
\be
\Pi_a + (p\cdot q) \Pi_b= -\Pi^{CT}((p+q)^2),
~~ (p\cdot q)\Pi_d= \Pi^{CT}(p^2) -\Pi^{CT}((p+q)^2).
\label{relcorr}
\ee
These constraints must be obeyed by the
OPE result, or any other calculation of the correlation function.
Now let us assume that we have calculated only the invariant
amplitude for the $p_\mu q_\nu$ structure in Eq.~(\ref{decompPi}). 
Without knowing 
the result for $\Pi_a$ and $\Pi_d$  
we can simply use the relations (\ref{relcorr}) and rewrite
the initial correlator as 
\bea
\label{decomPi1}
\Pi_{\nu} = e\Big\{\Big((\epsilon\cdot p)q_\nu - 
\epsilon_\nu(p\cdot q)\Big) \,
\Pi_b - \epsilon_\nu\Pi^{CT}((p+q)^2) 
\nonumber
\\
+ \frac{(\epsilon\cdot p)}{(p\cdot q)} 
p_\nu\Big(\Pi^{CT}(p^2)- \Pi^{CT}((p+q)^2)\Big) +i\epsilon_{\mu\nu
\lambda\sigma}
\epsilon^\mu p^\lambda q^\sigma\Pi_V \Big\}\,. 
\eea
We emphasize that the amplitudes in this expression
are fixed as a result of a direct calculation.
But we note that the relation (\ref{relcorr}) can be used
in a more general  way (compare Eqs.~(\ref{relab}) and (\ref{arbi})),
to include contact terms proportional to  $\epsilon_\nu$ as well
as to $q_\nu$. In analogy to the previous considerations
we have anticipated here the choice corresponding to $\beta=0$, that
is the form (\ref{decomPi1}) without a contact term proportional to  
$q_\nu$.

The next step in the sum rule derivation is in writing down the 
dispersion relation by inserting in Eq.~(\ref{corr})
a  complete set of hadronic states with the $B$ meson quantum numbers,
\bea
\label{corr1}
\Pi_\nu (p,q) = \frac{\langle 0 \mid \bar{u}
\Gamma_\nu b \mid B\rangle_{F(q)}\langle B \mid m_b \bar{b}i\gamma_5 u\}
\mid 0\rangle}{m_B^2-(p+q)^2}+ ...\,. 
\nonumber
\\
=\frac{e\epsilon^\mu T^{(B)}_{\mu\nu}f_Bm_B^2}{m_B^2-(p+q)^2}+ ...\,. 
\eea
We will also need the dispersion relation for the correlator 
$\Pi^{CT}$ which reads:
\bea
 \label{corr2}
\Pi^{CT}_\nu(p+q) = \frac{\langle 0 \mid \bar{u}
\Gamma_\nu b \mid B \rangle \langle B \mid m_b \bar{b}i\gamma_5 u\}
\mid 0\rangle}{m_B^2-(p+q)^2}+ ...\,. 
\nonumber
\\
=\frac{-f_B^2m_B^2}{m_B^2-(p+q)^2}(p+q)_\nu+ ...\,. 
\eea
In the above we retained the ground-state $B$ meson 
terms denoting the contribution of excited and continuum states
by ellipses. The latter states are not important for our analysis 
because their contributions one by one obey the same symmetry properties
as the $B$ meson term. The matrix element of the photon emission 
from $B$ meson entering  Eq.~(\ref{corr1}) 
can now be calculated matching the dispersion relation (\ref{corr1})
with the result of OPE for the correlation function $\Pi_\nu$. 
Simultaneously, the dispersion 
relation (\ref{corr2}) for $\Pi^{CT}$ yields 
the usual two-point QCD sum rule result for $f_B$.
We skip important points of the sum-rule procedure 
(use of quark-hadron duality,
continuum subtraction and Borel transformation) which are explained 
in detail in the literature (see e.g. \cite{CK}).
In particular, after the Borel transformation in the relevant 
variable $(p+q)^2$ the term proportional to $\Pi^{CT}(p^2)$ 
in Eq.~(\ref{decomPi1}) vanishes. 
The most important circumstance for our analysis is the fact 
that each invariant amplitude in the expansion (\ref{decomPi1})
yields, via the dispersion relation, a separate  sum-rule relation 
for the corresponding invariant amplitude in the matrix element. 
Thus, as a result of sum rule calculation and taking into 
account Eq.~(\ref{corr2})
one obtains 
\bea
\epsilon^\mu T^{(B)}_{\mu\nu}\,
= \left(\epsilon_\nu(p\cdot q) -(\epsilon\cdot p) q_\nu
\right)\,i\bar{F}^{(B)}_A
+i\bar{f}_B \epsilon_\nu 
\nonumber
\\
+i\frac{(\epsilon\cdot p) p_\nu}{(p\cdot q)}\bar{f}_B
+\epsilon_{\nu\mu \lambda\sigma}\epsilon^\mu p^\lambda q^\sigma
\bar{F}_{V}^{(B)}\,,
\label{decompim}
\eea
where $\bar{F}^{(B)}_A$, $\bar{F}^{(B)}_V$ and $\bar{f}_B$ are 
the hadronic amplitudes calculated from the QCD sum rules 
for $\Pi_b$ , $\Pi_V$ and $\Pi^{CT}$ respectively.   
The above equation exactly reproduces  the structure of Eq.~(\ref{decomp2}) 
together with the contact term.

Thus, we have shown that the calculational procedure 
which starts from the structure $p_\mu q_\nu$ in the
correlator yields the proper gauge-invariant term and
a contact term with structure $g_{\mu \nu}$. This latter is an inseparable 
part of the matrix element and cancels the (contact) term 
generated by the final state emission.

Finally, let us illustrate the above analysis by taking 
one of the short-distance contributions to the correlation
function, namely the quark-condensate term. 
This contribution
is not very important numerically in the final sum rule 
but is simply calculated and therefore convenient as a study case.
Physically, it corresponds to the photon emission at short distances 
from the virtual $b$ quark line whereas light $u$ quarks are soft 
and approximated by a local quark condensate.
The expression for this part of the correlation function 
is easily obtained by taking in Eq.~(\ref{pimunu})
$j_\mu= e_b \bar{b}\gamma_\mu b$, 
contracting the $b$ quark fields  
into free-quark propagators, and replacing the $u$ quark fields by 
the vacuum condensate $\langle \bar{u} u \rangle$ (further details
see in \cite{KSW}). The result reads:
\bea 
\Pi_{\mu\nu}^{\langle \bar{u} u \rangle}=
ie_b\frac{\langle\bar{u} u \rangle m_b}{(p^2-m_b^2)((p+q)^2-m_b^2)}&&
\nonumber
\\
\times \left[g_{\mu\nu}( m_b^2-p^2 -(p\cdot q) ) + p_\mu q_\nu + q_\mu p_\nu
+2p_\mu p_\nu +i\epsilon _{\mu\nu\alpha\beta}p^\alpha q_\beta\right] &&
\label{corrqq}
\eea
It is easy to check that the Ward identity is indeed valid and that 
the contact term is equal to the quark condensate contribution
to the two-point correlator:
\be
\Pi^{CT\langle \bar{u} u \rangle}(r^2)=ie_b\frac{ \langle \bar{u} u
\rangle m_b}{ r^2-m_b^2}\,.
\label{ctqq}
\ee
Using the latter expression we can rewrite Eq.~(\ref{corrqq}) in 
the form of Eq.~(\ref{decomPi1}) where 
\be
\Pi_b^{\langle \bar{u} u \rangle}=\Pi_V^{\langle \bar{u} u \rangle}
=ie_b\frac{\langle \bar{u} u \rangle m_b}{ (p^2-m_b^2)((p+q)^2-m_b^2}\,,
\ee
which agrees with the expression given in \cite{KSW}. Thus we
can uniquely identify the relevant form factor; again it is the
coefficient of the $(\epsilon\cdot p)q_\nu$ structure.
We have checked that the more complicated short-distance
contributions of perturbative diagrams 
to the sum rule calculated in \cite{KSW} 
are also in accordance with this procedure.

To summarize: In this paper we have identified a procedure based on
simple Ward identities to extract the correct form factors in the
calculation of the weak annihilation contribution to 
the radiative decays of the form $B \to V \gamma$.
\section{Acknowledgments}

We thank our collaborator G. Stoll for numerous
discussions from which this work emerged. We are grateful to
V.~M.~Braun for useful comments. The work of A.K. is supported 
by  BMBF (Bundesministerium f\"ur Bildung und Forschung); 
D.W. was partially supported by Schweizerischer Nationalfonds.

{\bf Note added}: 
After this paper was finished we became aware of 
the  recent work \cite{Riaz} where 
Ward identities and symmetry relations 
are applied but where vector-dominance is used.
This is a special model and allows only limited statements, unlike the 
general QCD 
picture employed here. The conclusion reached in \cite{Riaz} seems to
be in disagreement with the general result we present.


\end{document}